\documentclass{llncs}
\usepackage{amstext,amsgen,latexsym}
\usepackage{amstext,amssymb,amsfonts,latexsym}
\usepackage{theorem}

 \newcommand{\real}{\mathbb{R}}
 \newcommand{\nat}{\mathbb{N}}

 \newcommand{\complex}{\mathbb{C}}

 \newcommand{\prob}{{\mathrm{Prob}}}

 \newcommand{\ie}{\textrm{i.e.},\hspace*{2mm}}
 \newcommand{\eg}{\textrm{e.g.},\hspace*{2mm}}
 
 \newcommand{\etalc}{\textrm{et al.}}

 \newcommand{\HH}{{\cal H}}
 \newcommand{\KK}{{\cal K}}
 \newcommand{\SSS}{{\cal S}}
 \newcommand{\PP}{{\cal P}}

 \newcommand{\p}{\mathrm{{\bf P}}}

 \newcommand{\comb}[2]{{\tiny \left( \begin{array}{c} #1 \\%
      #2 \end{array}\right)}}

\theoremstyle{plain}
\theoremheaderfont{\bfseries}
 \newcommand{\pair}[1]{\langle #1 \rangle}
 \newcommand{\qubit}[1]{| #1 \rangle}
 \newcommand{\bra}[1]{\langle #1 |}
 \newcommand{\ket}[1]{| #1 \rangle}
 \newcommand{\braket}[2]{\langle #1 | #2 \rangle}
 \newcommand{\ketbra}[2]{| #1 \rangle\langle #2 |}
 \newcommand{\trnorm}[1]{\| #1 \|_{\mathrm{tr}}}
 \newcommand{\cc}{\mathrm{C}}
 \newcommand{\qca}{\mathrm{QCA}}
 
 \newcommand{\sqcd}{\mathrm{sQCD}}

\newcommand{\ignore}[1]{}

\begin{document}
\title{Computational Complexity Measures of \\
Multipartite Quantum Entanglement}
\subtitle{(Extented Abstract)}
\titlerunning{Quantum Entanglement}
\author{\sc Tomoyuki Yamakami}
\authorrunning{\sc Tomoyuki Yamakami}
\institute{School of Information Technology 
and Engineering \\ 
University of Ottawa, Ottawa, Canada K1N 6N5}
\maketitle

\begin{abstract}
We shed new light on entanglement measures in multipartite quantum
systems by taking a computational-complexity approach toward
quantifying quantum entanglement with two familiar
notions---{approximability} and distinguishability.  Built upon the
formal treatment of partial separability, we measure the complexity of
an entangled quantum state by determining (i) how hard to approximate
it from a fixed classical state and (ii) how hard to distinguish it
from all partially separable states.  We further consider the
Kolmogorovian-style descriptive complexity of approximation and
distinction of partial entanglement.
\end{abstract}

\section{Computational Aspects of Quantum Entanglement}
\label{sec:introduction}

Entanglement is one of the most puzzling notions in the theory of
quantum information and computation. A typical example of an entangled
quantum state is the Bell state (or the EPR pair)
$(\qubit{00}+\qubit{11})/\sqrt{2}$, which played a major role in, \eg
superdense coding \cite{BW92} and quantum teleportation schemes
\cite{BBC+93}.  Entanglement can be viewed as a physical resource and
therefore can be quantified.  Today, bipartite pure state entanglement
is well-understood with information-theoretical notions of
entanglement measures (see the survey \cite{Hor01}).

These measures, nevertheless, do not address computational aspects of
the complexity of entangled quantum states. For example, although the
Bell state is maximally entangled, it is computationally constructed
from the simple classical state $\qubit{00}$ by an application of the
Hadamard and the Controlled-NOT operators. Thus, if the third party
gives us a quantum state which is either the Bell state or any
separable state, then one can easily tell with reasonable confidence
whether the given state is truly the Bell state by reversing the
computation since the minimal trace distance between the Bell state
and separable states is at least $1/2$. This simple fact makes the
aforementioned information-theoretical measures unsatisfactory from a
computational point of view. We thus need different types of measures
to quantify multipartite quantum entanglement.

We first need to lay down a mathematical framework for multipartite
quantum entanglement and develop a useful terminology to describe a
nested structure of entangled quantum states. In this paper, we mainly
focus on pure quantum states in the Hilbert space $\complex^{2^n}$ of
dimension $2^n$. Such a state is called, analogous to a classical
string, a {\em quantum string} (or qustring, for short) {\em of length
$n$}. Any qustring of length $n$ is expressed in terms of the standard
basis $\{\qubit{s}\}_{s\in\{0,1\}^n}$. Given a qustring
$\qubit{\phi}$, let $\ell(\qubit{\phi})$ denote its length. By
$\Phi_n$ we denote the collection of all qustrings of length $n$ and
set $\Phi_{\infty}$ to be $\bigcup_{n\in\nat^{+}}\Phi_{n}$, where
$\nat^{+}=\nat-\{0\}$.  Ensembles (or series) of qustrings of
(possibly) different lengths are of particular interest. We use
families of {\em quantum circuits} \cite{Deu89,Yao93} as a
mathematical model of quantum-mechanical computation. A quantum
circuit has input qubits and (possibly) ancilla qubits, where all
ancilla qubits are always set to $\qubit{0}$ at the beginning of
computation. We fix a finite universal set of quantum gates, including
the identity and the NOT gate. As a special terminology, we say that a
property $\PP(n)$ holds for {\em almost all} (or {\em any sufficiently
large}) $n$ in $\nat$ if the set $\{x\in \nat\mid \mbox{ $\PP(x)$ does
not hold }\}$ is finite.  All logarithms are conventionally taken to
base two.

\section{Separability Index and Separability Distance}

We begin with a technical tool to identify the entanglement structure
of an arbitrary quantum state residing in a multipartite quantum
system.  In a bipartite quantum system, any separable state can be
expressed as a tensor product $\qubit{\phi}\otimes\qubit{\psi}$ of two
qubits $\qubit{\phi}$ and $\qubit{\psi}$ and thus, any other state has
its two qubits entangled with a physical correlation or ``bonding.''
In a multipartite quantum system, however, all ``separable'' states
may not have such a simple tensor-product form.  Rather, various
correlations of entangled qubits may be {\em nested}---or intertwined
over different groups of entangled qubits.  For example, consider the
qustring $\qubit{\psi_{2n}}=2^{-n/2}\sum_{x\in\{0,1\}^n}\qubit{xx}$ of
length $2n$. For each $i\in\{1,2,\ldots,n\}$, the $i$th qubit and the
$n+i$th qubit in $\qubit{\psi_{2n}}$ are entangled.  The reordering of
each qubit, nevertheless, unwinds its nested correlations and sorts
all the qubits in the blockwise tensor product form
$\qubit{\psi_{2n}'}=(\frac{1}{\sqrt{2}}(\qubit{00}+\qubit{11}))^{\otimes
n}$. Although $\qubit{\psi_{2n}}$ and $\qubit{\psi_{2n}'}$ are
different inputs for a quantum circuit, such a reordering is done at
the cost of additional $O(n)$ quantum gates. Thus, the number of those
blocks represents the ``degree'' of the separability of the given
qustring. Our first step is to introduce the appropriate terminology
that can describe this ``nested'' bonding structure of a qustring.

We introduce the structural notion, {\em separability index}, which
indicates the maximal number of entangled ``blocks'' that build up a
target qustring of a multipartite quantum system. See \cite{VPJK97}
also for multipartite separability.

\begin{definition}
1. For any two qustrings $\qubit{\phi}$ and $\qubit{\psi}$ of length
$n$, we say that $\qubit{\phi}$ is {\em isotopic} to $\qubit{\psi}$
via a permutation\footnote{Let $\sigma$ be any permutation on
$\{1,2,\ldots,n\}$ and let $\qubit{\phi}$ be any qustring of length
$n$. The notation $\sigma(\qubit{\phi})$ denotes the qustring that
results from permuting its qubits by $\sigma$; that is,
$\sigma(\qubit{\phi})=
\sum_{\vec{x}}\alpha_{\vec{x}}\qubit{x_{\sigma(1)}x_{\sigma(2)}\cdots
x_{\sigma(n)}}$ if
$\qubit{\phi}=\sum_{\vec{x}}\alpha_{\vec{x}}\qubit{x_1x_2\cdots x_n}$,
where $\vec{x}=x_1x_2\cdots x_n$ runs over all binary strings of
length $n$.} $\sigma$ on $\{1,2,\ldots,n\}$ if
$\sigma(\qubit{\phi})=\qubit{\psi}$.

2. A qustring $\qubit{\phi}$ of length $n$ is called {\em
$k$-separable} if $\qubit{\phi}$ is isotopic to
$\qubit{\phi_1}\otimes\qubit{\phi_2}\otimes\cdots\otimes
\qubit{\phi_k}$ via a certain permutation $\sigma$ on
$\{1,2,\ldots,n\}$ for a certain $k$-tuple
$(\qubit{\phi_1},\qubit{\phi_2},\ldots,\qubit{\phi_k})$ of qustrings
of length $\geq1$. This permutation $\sigma$ is said to {\em achieve
the $k$-separability of $\qubit{\phi}$} and the isotopic state
$\qubit{\phi_1}\otimes\qubit{\phi_2}\otimes\cdots\otimes
\qubit{\phi_k}$ is said to have a {\em $k$-unnested form}. The series
$\vec{m}=
(\ell(\qubit{\phi_1}),\ell(\qubit{\phi_2}),\ldots,\ell(\qubit{\phi_k}))$
is called a {\em $k$-sectioning} of $\qubit{\phi}$ by $\sigma$.

3. The {\em separability index} of $\qubit{\phi}$, denoted
$sind(\qubit{\phi})$, is the maximal integer $k$ with $1\leq k\leq n$
such that $\qubit{\phi}$ is $k$-separable.
\end{definition}

For any indices $n,k\in\nat^{+}$ with $k\leq n$, let $QS_{n,k}$ denote
the set of all qustrings of length $n$ that have separability index
$k$.

For clarity, we re-define the terms ``entanglement'' and
``separability'' using the separability indices. These terms are
different from the conventional ones.

\begin{definition}
For any qustring $\qubit{\phi}$ of length $n$, $\qubit{\phi}$ is {\em
fully entangled} if its separability index equals $1$ and
$\qubit{\phi}$ is {\em fully separable} if it has separability index
$n$. For technicality, we call $\qubit{\phi}$ {\em partially
entangled} if it is of separability index $\leq n-1$. Similarly, a
{\em partially separable} qustring is a qustring with separability
index $\geq2$.
\end{definition}
 
We assume the existence of a {\em quantum source of information};
namely, a certain physical process that produces a stream of quantum
systems (\ie qustrings) of (possibly) different lengths. Such a
quantum source generates an ensemble (or a series) of qustrings. Of
such ensembles, we are particularly interested in the ensembles of
partially entangled qustrings. For convenience, we call them {\em
entanglement ensembles}.

\begin{definition}
Let $\ell$ be any strictly increasing function from $\nat$ to
$\nat$. A series $\Xi=\{\qubit{\xi_n}\}_{n\in\nat}$ is called an {\em
entanglement ensemble with size factor $\ell$} if, for every index
$n\in\nat$, $\qubit{\xi_n}$ is a partially entangled qustring of
length $\ell(n)$.
\end{definition}

How close is a fully entangled state to its nearest partially
separable state?  Consider the fully entangled qustring
$\qubit{\phi_n}=(\qubit{0^n}+\qubit{1^n})/\sqrt{2}$ for any
$n\in\nat$.  For comparison, let $\qubit{\psi}$ be any partially
separable qustring of length $n$.  By a simple calculation, the
$L_2$-norm distance $\|\qubit{\phi_n} - \qubit{\psi}\|$ is shown to be
at least $\sqrt{2-\sqrt{2}}$.  The Bures metric
$B(\qubit{\phi_n},\qubit{\psi})=2(1-F(\qubit{\phi_n},\qubit{\psi}))$,
where $F$ is the fidelity,\footnote{There are two different
definitions in the literature. Following \cite{NC00}, we define the
fidelity of two density operators $\rho$ and $\tau$ as
$F(\rho,\tau)=Tr(\sqrt{\sqrt{\rho}\tau\sqrt{\rho}})$.} is at least
$2-\sqrt{2}$ since we have $F(\qubit{\phi_n},\qubit{\psi})\leq
1/\sqrt{2}$ using the equality
$F(\qubit{\phi_n},\qubit{\psi})=|\braket{\phi_n}{\psi}|$.  The trace
distance\footnote{The trace norm of a linear operator $X$ is defined
as $\trnorm{X}=\frac{1}{2}Tr(\sqrt{X^{\dagger}X})$ \cite{NC00}.}
$\trnorm{\ket{\phi_n}\bra{\phi_n} - \ket{\psi}\bra{\psi}}$ is bounded
below by $1/2$ using the inequality
$1-F(\qubit{\phi_n},\qubit{\psi})^2\leq
\trnorm{\ket{\phi_n}\bra{\phi_n} - \ket{\psi}\bra{\psi}}$ and the
above bound for the fidelity.

This example motivates us to introduce the following notion of
``closeness'' similar to \cite{VPRK97} using the trace norm. Note that
the choice of a distance measure is not essential for our study.

\begin{definition}\label{def:close}
Let $k,n\in\nat^{+}$, $\delta\in[0,1]$, and let $\qubit{\xi}$ be any
qustring of length $n$.

1. The {\em $k$-separability distance} of $\qubit{\xi}$, denoted
$sdis_{k}(\qubit{\xi})$, is the infimum of $\trnorm{\ket{\xi}\bra{\xi}
-\ket{\phi}\bra{\phi}}$ over all $k$-separable qustrings
$\qubit{\phi}$ of length $n$.

2. A qustring $\qubit{\xi}$ is said to be {\em $(k,\delta)$-close to
separable states} if $sdis_{k}(\qubit{\xi})\leq\delta$. Otherwise,
$\qubit{\xi}$ is {\em $(k,\delta)$-far from separable states}.

3. Let $k$ be any function from $\nat$ to $\nat^{+}$ and let $\delta$
be any function from $\nat$ to $[0,1]$.  An ensemble
$\Xi=\{\qubit{\xi_n}\}_{n\in\nat}$ of qustrings is {\em
$(k,\delta)$-close} ({\em infinitely-often $(k,\delta)$-close}, resp.)
{\em to separable states} if $\qubit{\xi_n}$ is
$(k(n),\delta(n))$-close to separable states for almost all $n\in\nat$
(for infinitely many $n\in\nat$, resp.).  We say that $\Xi$ is {\em
$(k,\delta)$-far} ({\em infinitely-often $(k,\delta)$-far}, resp.)
{\em from separable states} if $\qubit{\xi_n}$ is
$(k(n),\delta(n))$-far from separable states for almost all $n\in\nat$
(for infinitely many $n\in\nat$, resp.).
\end{definition}

Notice that $sdis_{k}(\qubit{\xi})=0$ if $\qubit{\xi}$ is
$k$-separable. Moreover, the $k$-separability distance is invariant to
permutation; namely, $sdis_{k}(\sigma(\qubit{\xi}))=
sdis_{k}(\qubit{\xi})$ for any permutation $\sigma$. The previous
example shows that the entanglement ensemble
$\{(\qubit{0^n}+\qubit{1^n})/\sqrt{2}\}_{n\in\nat}$ are
$(2,1/2-\epsilon)$-far from separable states for any constant
$\epsilon>0$. Our measure also has a connection to the geometric
measure (see \cite{WG03} for a review).

The notion of von Neumann entropy\footnote{The von Neumann entropy
$S(\rho)$ of a density operator $\rho$ is $-Tr(\rho\log\rho)$, where
the logarithm is taken to base 2. See, \eg \cite{NC00}.} has been
proven to be useful for the characterization of entanglement of
bipartite pure quantum states. The von Neumann entropy measures the
{\em mixedness} of a mixed quantum state.  Let $\qubit{\psi}$ be any
qustring of length $n$. For each $i\in\{1,\ldots,n\}$, let $\HH_{\geq
i}$ denote the Hilbert space corresponding to the last $n-i+1$st
qubits of $\qubit{\psi}$.  Consider the set $\SSS= \{S(Tr_{\HH_{\geq
i}}(\ketbra{\psi}{\psi}))\mid i=2,3,\ldots,n\}$, where $Tr_{\HH_{\geq
i}}$ is the trace-out operator\footnote{For any bipartite quantum
system $\HH\otimes\KK$, the {\em trace-out operator} (or {\em partial
trace}) $Tr_{\KK}$ is the mapping defined by $Tr_{\KK}(\rho) =
\sum_{j=1}^{n}(I\otimes\bra{e_j})\rho(I\otimes\ket{e_j})$ for any
density operator $\rho$ of $\HH\otimes\KK$, where
$\{\qubit{e_1},\ldots,\qubit{e_n}\}$ is any fixed orthonormal basis of
$\KK$.}.  We define the {\em average entropy} of $\qubit{\psi}$ as
$E(\ketbra{\psi}{\psi})=\frac{1}{n-1}\sum_{i=2}^{n}S(Tr_{\HH_{\geq
i}}(\ketbra{\phi}{\phi}))$.  The following lemma then holds.

\begin{lemma}\label{lemma:vonNeumann}
Let $n\in\nat^{+}$, $\qubit{\xi}\in\Phi_{n}$, and
$k\in\{2,3,\ldots,n\}$.  If $sdis_{k}(\qubit{\xi})\leq 1/e$, then
$\min_{\qubit{\phi}}\{|E(\ketbra{\xi}{\xi})- E(\ketbra{\phi}{\phi})|\}
\leq sdis_{k}(\qubit{\xi})(n - \log{sdis_{k}(\qubit{\xi})})$, where
the minimization is taken over all $k$-separable qustrings in
$\Phi_{n}$.
\end{lemma}

For Lemma \ref{lemma:vonNeumann}, note that $|E(\ketbra{\xi}{\xi})-
E(\ketbra{\phi}{\phi})|\leq
\frac{1}{n-1}\sum_{i=2}^{n}|S(Tr_{\HH_{\geq i}}(\ketbra{\xi}{\xi})) -
S(Tr_{\HH_{\geq i}}(\ketbra{\phi}{\phi}))|$. By the Fanne inequality
(see, \eg \cite{NC00}), the difference \linebreak $|S(Tr_{\HH_{\geq
i}}(\ketbra{\xi}{\xi})) - S(Tr_{\HH_{\geq i}}(\ketbra{\phi}{\phi}))|$
is at most $\trnorm{Tr_{\HH_{\geq i}}(\ketbra{\xi}{\xi}) -
Tr_{\HH_{\geq i}}(\ketbra{\phi}{\phi})} \cdot \log{2^{i-1}} +
\eta(\trnorm{Tr_{\HH_{\geq i}}(\ketbra{\xi}{\xi}) - Tr_{\HH_{\geq
i}}(\ketbra{\phi}{\phi})})$, which is bounded by
$sdis_{k}(\qubit{\xi})[n-\log{sdis_{k}(\qubit{\xi})}]$, where
$\eta(\gamma)=-\gamma\log\gamma$ for $\gamma>0$.

\section{Entanglement Distinguishability}\label{sec:distinguishability}

We measure the complexity of each entangled state $\qubit{\phi}$ by
determining how hard it is to distinguish $\qubit{\phi}$ from all
$k$-separable states. Earlier, Vedral \etalc~\cite{VPJK97} recognized
the importance of distinguishability for quantifying
entanglement. Fuchs and van de Graaf \cite{FG99} took a cryptographic
approach to quantum state distinguishing problems and briefly
discussed computational indistinguishability of quantum states.

Cryptography has utilized the notion of ``distinguishers'' as, \eg an
adversary to a pseudorandom generator. Such a distinguisher is
designed to distinguish between two different distributions of strings
of fixed length with reasonable confidence. Since a quantum state can
be viewed as an extension of a classical distribution, we can
naturally adapt this cryptographic concept into a quantum context. For
a quantum circuit $C$ and a density operator $\rho$, the notation
$C(\rho)$, ignoring ancilla qubits, stands for the random variable
describing the measured output bit of $C$ on input $\rho$. However,
for a qustring $\qubit{\phi}$, $C\qubit{\phi}$ denotes the quantum
state that results from $\qubit{\phi}$ by an application of $C$.

\begin{definition}
Let $\epsilon\in[0,1]$ and let $\rho$ and $\tau$ be any two density
operators of the same dimension. We say that a quantum circuit $C$
{\em $\epsilon$-distinguishes} between $\rho$ and $\tau$ if
$|\prob_{C}[C(\rho)=1] - \prob_{C}[C(\tau)=1]|\geq\epsilon$. This
circuit $C$ is called an {\em $\epsilon$-distinguisher} of $\rho$ and
$\tau$.
\end{definition}

Now, we introduce a special type of distinguisher, which distinguishes
a given ensemble of partially entangled qustrings from $k$-separable
states using only polynomially-many quantum gates.  Let $\qubit{\phi}$
be any $k$-separable qustring of length $n$ that is isotopic to the
state
$\qubit{\phi_1}\otimes\qubit{\phi_2}\otimes\cdots\otimes\qubit{\phi_k}$
via a permutation $\sigma$.  Let $\vec{m}=
(\ell(\qubit{\phi_1}),\ldots,\ell(\qubit{\phi_k}))$ be its
$k$-sectioning. For notational convenience, we write $1^{\vec{m}}$ for
$1^{\ell(\qubit{\phi_1})}01^{\ell(\qubit{\phi_2})}0\cdots
1^{\ell(\qubit{\phi_k})}0$ whose length is exactly $n+k$. Let
$1^{\sigma}$ be $1^{\sigma(1)}01^{\sigma(2)}0\cdots 1^{\sigma(n)}0$ of
length $n^2/2+3n/2$. Moreover, we write $1^{\sigma,\vec{m}}$ for
$1^{\sigma}01^{\vec{m}}0$. Note that the length of
$1^{\sigma,\vec{m}}$ is $n^2/2+5n/2+k+2$.

\begin{definition}\label{def:distinguishable}
Let $k$ be any function from $\nat$ to $\nat-\{0,1\}$ and $\epsilon$
be any function from $\nat$ to $[0,1]$. Let $\ell$ and $s$ be any
functions from $\nat$ to $\nat$. Assume that $\ell$ is strictly
increasing.  Let $\Xi=\{\qubit{\xi_n}\}_{n\in\nat}$ be an ensemble of
qustrings with size factor $\ell$.

1. A family $\{D_{n}\}_{n\in\nat}$ of quantum circuits with
$\ell(n)^2/2+7\ell(n)/2+k(n)+2$ input qubits and (possibly) ancilla
qubits is called a {\em non-uniform entanglement
$(k,\epsilon)$-distinguisher} ({\em non-uniform infinitely-often
entanglement $(k,\epsilon)$-distinguisher}, resp.) {\em of $\Xi$} if,
for almost all $n$'s (for infinitely many $n\in\nat$, resp.), $D_n$
$\epsilon(n)$-distinguishes between
$\qubit{1^{\sigma,\vec{m}}}\qubit{\xi_n}$ and
$\qubit{1^{\sigma,\vec{m}}}\qubit{\phi}$ for any $k$-separable
qustring $\qubit{\phi}$ of length $\ell(n)$ and any permutation
$\sigma$ that achieves the $k$-separability of $\qubit{\phi}$ with
$k(n)$-sectioning $\vec{m}$. In particular, if we want to emphasize a
pair $(\sigma,\vec{m})$, we call $D_n$ a {\em non-uniform
(infinitely-often) entanglement $\epsilon$-distinguisher with respect
to $(\sigma,\vec{m})$}.

2. The ensemble $\Xi$ is called {\em non-uniformly
$(k,\epsilon,s)$-distinguishable from separable states} if there is a
non-uniform entanglement $(k,\epsilon)$-distinguisher of $\Xi$ that
has size\footnote{The {\em size} of a quantum circuit is the total
number of quantum gates in it.} at most $s(n)$. In contrast, $\Xi$ is
{\em non-uniformly $(k,\epsilon,s)$-indistinguishable from separable
states} if there is no $s$-size non-uniform infinitely-often
entanglement $(k,\epsilon)$-distinguisher of $\Xi$. In case where $s$
is a polynomial, we simply say that $\Xi$ is {\em non-uniformly
$(k,\epsilon)$-distinguishable from separable states} and {\em
non-uniformly $(k,\epsilon)$-indistinguishable from separable states},
respectively. Similarly, we can define the infinitely-often version of
distinguishability and indistinguishability. For readability, we often
drop the word ``non-uniform'' if it is clear from the context.
\end{definition}

We can also define a ``uniform'' entanglement distinguisher using a
$\p$-uniform family of quantum circuits (or equivalently, a multi-tape
quantum Turing machine \cite{BV97,Yam99}).

Obviously, any ensemble of $k$-separable qustrings is
$(k,\epsilon)$-indistinguishable from separable states for any
$\epsilon\geq0$. The following lemma is an immediate consequence of
Definition \ref{def:distinguishable}.

\begin{lemma}\label{lemma:property-dist}
Let $\Xi=\{\qubit{\xi_n}\}_{n\in\nat}$ be any ensemble of qustrings
with size factor $\ell$.

1. Let $k,k'$ be any functions from $\nat$ to $\nat-\{0,1\}$, let
$\epsilon,\epsilon'$ be any functions from $\nat$ to $[0,1]$, and let
$s,s'$ be any functions from $\nat$ to $\nat$. Assume that $k'$,
$\epsilon$ and $s'$ majorize\footnote{For any two functions $f,g$ from
$\nat$ to $\real$, we say that $f$ {\em majorizes} $g$ if $g(n)\leq
f(n)$ for every $n\in\nat$.} $k$, $\epsilon'$, and $s$,
respectively. If $\Xi$ is (infinitely-often)
$(k,\epsilon,s)$-distinguishable from separable states, then $\Xi$ is
also (infinitely-often) $(k',\epsilon',s')$-distinguishable from
separable states.

2. Let $\vec{\sigma}=\{\sigma_n\}_{n\in\nat}$ be any family of
   permutations $\sigma_n$ on $\{1,2,\ldots,\ell(n)\}$ for each
   $n$. Define $\vec{\sigma}(\Xi)
   =\{\sigma_n(\qubit{\xi_n})\}_{n\in\nat}$. If $\Xi$ is
   (infinitely-often) $(k,\epsilon,s)$-distinguishable from separable
   states, then $\vec{\sigma}(\Xi)$ is (infinitely-often)
   $(k,\epsilon,s(n)+O(n))$-distinguishable from separable states.
\end{lemma}

Of course, there are entangled states that no quantum circuit can
distinguish from separable states. For instance, if two qustrings are
close to each other, then no polynomial-size quantum circuit can tell
their difference.  In what follows, we show that any entangled state
close to separable states is indistinguishable.

\begin{proposition}\label{prop:close-indistinguish}
Let $k$ be any function from $\nat$ to $\nat^{+}$ and let $\ell$ be
any function from $\nat$ to $\nat$. Any entanglement ensemble
$\Xi=\{\qubit{\xi_i}\}_{i\in\nat}$ is
$(k(n),sdis_{k(n)}(\qubit{\xi_n})+\delta)$-indistinguishable from
separable states for any constant $\delta>0$.
\end{proposition}

Proposition \ref{prop:close-indistinguish} is proven by the inequality
$|\prob_{C}[C(\qubit{1^{\sigma,\vec{m}}}\qubit{\xi_n})=1] - 
\linebreak
\prob_{C}[C(\qubit{1^{\sigma,\vec{m}}}\qubit{\phi})=1]|\leq
\trnorm{\ketbra{\xi_n}{\xi_n} - \ketbra{\phi}{\phi}}$ for any
$k$-separable state $\qubit{\phi}$, which follows from the fact that
$\trnorm{\rho - \sigma} = \max_{P}\{Tr(P(\rho - \sigma))\}$, where the
maximization is taken over all positive semidefinite
contractive\footnote{A square matrix $A$ is {\em contractive} if
$\|A\|\leq1$, where
$\|A\|=\sum_{\qubit{\phi}\neq0}\{\|A\qubit{\phi}\|/\|\qubit{\phi}\|\}$.}
matrices $P$.

We note that, for every qustring $\qubit{\xi}\in\Phi_{n}$, there
exists a positive operator-valued measure $W$ such that
$\max_{\qubit{\phi}}\{|\bra{\xi}W\ket{\xi} -
\bra{\phi}W\ket{\phi}|\}\geq sdis_{k}(\qubit{\xi})^2$, where the
maximization is taken over all $k$-separable qustrings in
$\Phi_{n}$. Such a $W$ is given, for example, as $W= I -
\ketbra{\xi}{\xi}$. Lemma \ref{lemma:construct-distinguish} will
present its special case.

How do we construct our distinguisher? A basic way is to combine all
distinguishers built with respect to different pairs of permutations
and sectionings. Suppose that we have $s$-size entanglement
distinguishers with respect to permutations $\sigma$ and
$k(n)$-sectionings $\vec{m}$ targeting the same entanglement ensemble
$\Xi$ with size factor $\ell(n)$. Although the number of such pairs
$(\sigma,\vec{m})$ may be nearly
$\ell(n)!\cdot\comb{\ell(n)-1}{k(n)}$, the following lemma shows that
it is possible to build a $O(s)$-size distinguisher that works for all
permutation-sectioning pairs.

\begin{lemma}\label{lemma:combining}
Let $\Xi$ be any entanglement ensemble with size factor $\ell$. Let
$s$ be any strictly increasing function from $\nat$ to $\nat$. If, for
every $n\in\nat$, every permutation $\sigma$ on
$\{1,\ldots,\ell(n)\}$, and every $k(n)$-sectioning $\vec{m}$, there
exists an $s(n)$-size $\epsilon$-distinguisher of $\Xi$ with respect
to $(\sigma,\vec{m})$, then there exists an $O(s(n)^c)$-size
$(k,\epsilon)$-distinguisher of $\Xi$, where $c$ is an absolute
positive constant.
\end{lemma}

\section{Entanglement Approximability}\label{sec:approximability}

What types of entangled states are easily distinguishable from
separable states? We first claim that any entangled state that is
computationally ``constructed'' from the classical state $\qubit{0^m}$
is distinguishable. The precise definition of constructibility is
given as follows.

\begin{definition}
Let $s$ be any function from $\nat$ to $\nat$. An ensemble
$\Xi=\{\qubit{\xi_n}\}_{n\in\nat}$ of qustrings with size factor
$\ell(n)$ is {\em non-uniformly $s$-size constructible} if there
exists a non-uniform family $\{C_n\}_{n\in\nat}$ of quantum circuits
of size at most $s(n)$ having $\ell(n)$ input qubits and no ancilla
qubit such that, for every $n$,
$C_n\qubit{0^{\ell(n)}}=\qubit{\xi_n}$, where $C_n\qubit{0^{\ell(n)}}$
denotes the qustring obtained after the computation of $C_n$ on input
$\qubit{0^{\ell(n)}}$. This family $\{C_n\}_{n\in\nat}$ is called a
{\em non-uniform $s$-size constructor} of $\Xi$.
\end{definition}

Consider a partially entangled qustring $\qubit{\xi}$ of length $n$
with $\delta=sdis_{k}(\qubit{\xi})>0$. If $\qubit{\xi}$ is
computationally constructed from $\qubit{0^{n}}$, then we can easily
determine whether a quantum state given from the third party is
exactly $\qubit{\xi}$ by reversing the construction process to test
whether it returns to $\qubit{0^n}$. This induces a distinguisher
$D$. This is seen as follows. For any $k$-separable state
$\qubit{\phi}$, we have $|\prob_{D}[D(\qubit{\xi})=1] -
\prob_{D}[D(\qubit{\phi})=1]|\geq \delta^2$ since
$\prob_{D}[D(\qubit{\phi})=1]=F(\qubit{\xi},\qubit{\phi})^2$, which is
bounded above by $1-\delta^2$. Therefore, we obtain:

\begin{lemma}\label{lemma:construct-distinguish}
Let $\ell$ and $s$ be any functions from $\nat$ to $\nat$ and $k$ be
any function from $\nat$ to $\nat^{+}$. Assume that $\ell$ is strictly
increasing. For any ensemble $\Xi=\{\qubit{\xi_n}\}_{n\in\nat}$ of
qustrings of size factor $\ell$, if $\Xi$ is $s$-size constructible,
then it is
$(k(n),sdis_{k(n)}(\qubit{\xi_n})^2,O(s(n)))$-distinguishable from
separable states.
\end{lemma}

Many fully entangled quantum states used in the literature are
polynomial-size constructible. For instance, the entanglement ensemble
$\{(\qubit{0^n}+\qubit{1^n})/\sqrt{2}\}_{n\in\nat}$ is $O(n)$-size
constructible and its $2$-separability distance is at least
$1/2$. Thus, it is $(2,1/4,O(n))$-distinguishable from separable
states.

We further relax the computability requirement for partially entangled
states. Below, we introduce quantum states that can be
``approximated'' rather than ``constructed.''

\begin{definition}
Let $s$ be any function from $\nat$ to $\nat$ and $\epsilon$ be any
function from $\nat$ to $[0,1]$. An ensemble
$\Xi=\{\qubit{\xi_n}\}_{n\in\nat}$ of qustrings with size factor
$\ell(n)$ is said to be {\em non-uniformly
$(\epsilon,s)$-approximable} ({\em non-uniformly infinitely-often
$(\epsilon,s)$-approximable}, resp.) if there exists a non-uniform
family $\{C_n\}_{n\in\nat}$ of quantum circuits of size at most $s(n)$
having $\ell(n)$ input qubits and $p(n)$ ancilla qubits ($p(n)\geq0$)
such that, for almost all $n\in\nat$ (for infinitely many $n\in\nat$,
resp.),
\[
\left\|Tr_{\HH_n}(C_n\ket{0^{p(n)+\ell(n)}}\bra{0^{p(n)+\ell(n)}}
C_n^{\dagger}) - \ket{\xi_n}\bra{\xi_n}\right\|_{\mathrm{tr}}\leq 
\epsilon(n),
\]
where $\HH_n$ refers to the Hilbert space corresponding to the $p(n)$
ancilla qubits of $C_n$. The family $\{C_n\}_{n\in\nat}$ is called a
{\em non-uniform $\epsilon$-approximator} ({\em non-uniform
infinitely-often $\epsilon$-approximator}, resp.) of $\Xi$. In
particular, if $\Xi$ is non-uniformly (infinitely-often)
$(\epsilon,s)$-approximable for a certain polynomial $s$, then we
simply say that $\Xi$ is {\em non-uniformly (infinitely-often)
$\epsilon$-approximable}.
\end{definition}

The ``uniform'' version of approximability can be defined using a
$\p$-uniform family of quantum circuits or a multi-tape quantum Turing
machine. As seen before, we drop the phrase ``non-uniform'' in the
above definition for simplicity unless otherwise stated. Clearly, any
$(\epsilon,s)$-constructible quantum state is
$(\epsilon,s)$-approximable.

The following lemma shows that any ensemble of qustrings has an
exponential-size approximator; however, there exists an ensemble that
is not approximated by any polynomial-size approximators.

\begin{lemma}\label{lemma:approximator}
1. Let $\epsilon$ be any function from $\nat$ to $(0,1]$. Any ensemble
of qustrings with size factor $n$ has a non-uniform
$(\epsilon,s)$-approximator, where $s(n)=
n^22^{n}\log^2{\frac{n^22^{2n}}{\epsilon(n)}}$.

2. For each constant $\epsilon>0$, there exists an entanglement
ensemble that is not $(\epsilon,n^{O(1)})$-approximable.
\end{lemma}

Lemma \ref{lemma:approximator}(1) follows from the Solovay-Kitaev
theorem (see \cite{NC00}). Lemma \ref{lemma:approximator}(2) uses the
result in \cite{Kni95} that there exists a quantum state that is not
approximated by any polynomial-size quantum circuits together with the
fact that there is always an entangled state close to each separable
state.

A role of approximators is to build distinguishers. We can show that
approximability implies distinguishability if the target entanglement
ensemble is far from separable states.

\begin{proposition}\label{prop:approximate-distinguish}
Let $k$ be any function from $\nat$ to $\nat-\{0,1\}$ and
$\epsilon,\delta$ be any functions from $\nat$ to $[0,1]$ such that
$\delta(n)>\epsilon(n)+\sqrt{\epsilon(n)}$ for all $n$. For any
(infinitely-often) $(\epsilon,s)$-approximable entanglement ensemble,
if it is $(k,\delta)$-far from separable states, then it is
(infinitely-often) $(k,\epsilon',O(s(n)))$-distinguishable from
separable states, where $\epsilon'(n) =
\frac{(\delta(n)-\epsilon(n))^2-\epsilon(n)}{2}$.
\end{proposition}

The proof of Proposition \ref{prop:approximate-distinguish} is based
on the fact that any $(\epsilon,s)$-approximable entanglement ensemble
$\Xi=\{\qubit{\xi_n}\}_{n\in\nat}$ can be distinguished from separable
states by use of the {\sf Controlled-SWAP} operator (see
\cite{BCWW01,KMY01}). Let $D_n$ be the circuit that runs an
$(\epsilon(n),s(n))$-approximator $C_n$ and then carries out the
C-SWAP procedure (first apply the Hadamard $H$ to the controlled bit
$\qubit{0}$, then {\sf Controlled-SWAT}, and finally $H$) and outputs
the complement of the controlled bit. Let $\qubit{\psi}$ be any
qustring of length $\ell(n)$. It follows that
$\prob_{D_n}[D_n(\qubit{\psi})=1] = 1/2+
Tr(\rho\ketbra{\psi}{\psi})/2$, where
$\rho=Tr_{\HH_n}(C_n\ketbra{0^m}{0^m}C_n^{\dagger})$ for some
appropriate $m$. On one hand, we have
$\prob_{D_n}[D_n(\qubit{\xi_n})=1]\geq 1-\epsilon(n)/2$. On the other
hand, if $\qubit{\psi}$ is $k(n)$-separable and $(k(n),\delta(n))$-far
from separable states, then
$\prob_{D_n}[D_n(\qubit{\xi_n})=1]<1-(\delta(n)-\epsilon(n))^2/2$. 
Therefore, $|\prob_{D_n}[D_n(\qubit{\xi_n})=1] -
\prob_{D_n}[D_n(\qubit{\psi})=1]|$ is greater than
$\epsilon'(n)$. Note that Proposition
\ref{prop:approximate-distinguish} also holds for the uniform case.

Recall from Proposition \ref{prop:close-indistinguish} that any
entanglement ensemble close to separable states is indistinguishable
from separable states. Conversely, we claim a general result that any
entangled state that is far from separable states has exponential-size
distinguishers by combining Proposition
\ref{prop:approximate-distinguish} with Lemma
\ref{lemma:approximator}(1) as well as the fact that
$n^22^{n}\log^2\frac{n^22^{2n}}{\epsilon}\in O(2^{2n})$.

\begin{corollary}
Let $k$ be any function from $\nat$ to $\nat-\{0,1\}$ and
$\epsilon,\delta$ be any functions from $\nat$ to $[0,1]$ with
$\delta(n)>\epsilon(n)+\sqrt{\epsilon(n)}$ for any $n$. Every
entanglement ensemble that is $(k,\delta)$-far from separable states
is $(k,\epsilon',O(2^{2n}))$-distinguishable from separable states,
where $\epsilon'(n)=\frac{(\delta(n)-\epsilon(n))^2-\epsilon(n)}{2}$.
\end{corollary}

Under the uniformity condition, we can show that distinguishability
does not always imply approximability. To see this, consider the
entanglement ensemble
$\Xi=\{(\qubit{0^n}+(-1)^{f(1^n)}\qubit{1^n})/\sqrt{2}\}_{n\in\nat}$,
where $f$ is any recursive function from $\{1\}^*$ to $\nat$, which is
not computable by any $\p$-uniform family of exponential-size Boolean
circuits. This $\Xi$ can be uniformly
$(n,1/\sqrt{2},n^{O(1)})$-distinguishable but not uniformly
$(1/\sqrt{2},n^{O(1)})$-approximable; otherwise, we can build from an
approximator of $\Xi$ a family of exponential-size Boolean circuits
that compute $f$. Therefore, we obtain:

\begin{proposition}\label{prop:uniform-separation}
There exists an entanglement ensemble of size factor $n$ that is
uniformly $(n,1/\sqrt{2},n^{O(1)})$-distinguishable from separable
states and not uniformly $(1/\sqrt{2},n^{O(1)})$-approximable.
\end{proposition}

\section{Descriptive Complexity of Entanglement}\label{sec:Kolmogorov}

The recent work of Vit{\'a}nyi \cite{Vit01} and Berthiaume
\etalc~\cite{BDL00} brought in the notion of quantum Kolmogorov
complexity to measure the descriptive (or algorithmic) complexity of
quantum states. In particular, Vit{\'a}nyi measured the minimal size
of a classical program that approximates a target quantum state. We
modify Vit{\'a}nyi's notion to accommodate the approximability of
partially entangled qustrings using quantum circuits of bounded size.

Let us fix an appropriate universal deterministic Turing machine
$M_{U}$ and let $\cc(x|y)$ denote the Kolmogorov complexity of $x$
conditional to $y$ with respect to $M_{U}$; that is, the minimal
nonnegative integer $|p|$ such that $p$ is a classical program that
produces $x$ from $y$ (\ie $M_{U}(p,y)=x$ in finite time). Abbreviate
$\cc(x|\lambda)$ as $\cc(x)$.  By identifying a quantum circuit $D$
with its encoding\footnote{The notation $\pair{D}$ for a quantum
circuit $D$ denotes a fixed effective encoding of $D$ such that the
size of this coding is not smaller than the number of gates in $D$.}
$\pair{D}$, we succinctly write $\cc(D)$ for $\cc(\pair{D})$.
 
\begin{definition}
Let $s$ be any function from $\nat$ to $\nat$ and let $\qubit{\xi}$ be
any qustring of length $n$. The {\em $s$-size bounded approximating
complexity} of $\qubit{\xi}$, denoted $\qca^{s}(\qubit{\xi})$, is the
infimum of $\cc(D) - \log{F(\qubit{\xi},\rho)^2}$ such that $D$ is a
quantum circuit of size at most $s(n)$ with $\ell$ inputs ($\ell\geq
n$) and $\rho= Tr_{\HH}(\ketbra{\phi}{\phi})$, where $\qubit{\phi}=
D\qubit{0^{\ell}}$ and $\HH$ is the Hilbert space associated with the
last $\ell-n$ qubits of $D$. Its conditional version
$\qca^{s}(\qubit{\xi}|\qubit{\zeta})$ is defined by
$\cc(D|\ell(\qubit{\zeta}))-\log{F(\ketbra{\xi}{\xi},\sigma)^2}$,
where $\qubit{\psi}= D\qubit{\zeta}\qubit{0^{\ell}}$ and
$\sigma=Tr_{\HH}(\ketbra{\psi}{\psi})$.
\end{definition}

More generally, we can define $\qca^{s}(\sigma)$ for any density
operator $\sigma$. Similar to \cite{Vit01}, $\qca^{s}(\qubit{\xi})$ is
bounded above by $2n +c$ for any $\qubit{\xi}\in\Phi_{n}$ (by
considering a quantum circuit $C$ that outputs $\qubit{x}$ satisfying
$F(\qubit{\xi},\qubit{x})^2\geq 2^{-n}$) if $s(n)\geq n$.

We prove in the following lemma that any uniformly approximable
entanglement ensemble has small approximating complexity. This lemma
comes from the inequality $\trnorm{\rho - \ketbra{\xi}{\xi}} \leq
\sqrt{1-F(\rho,\qubit{\xi})^2}$.

\begin{lemma}\label{lemma:upper-bound}
Let $s$ be any function from $\nat$ to $\nat$ and let $\epsilon$ be
any function from $\nat$ to $[0,1)$. Let
$\Xi=\{\qubit{\xi_n}\}_{n\in\nat}$ be any entanglement ensemble.  If
$\Xi$ is uniformly $(\epsilon,s)$-approximable, then there exists an
absolute constant $c\geq0$ such that
$\qca^{s}(\qubit{\xi_n}|\qubit{1^n}) \leq c - \log(1-\epsilon(n))$ for
all $n\in\nat$. In particular, if $\epsilon(n)$ is upper-bounded by a
certain constant, then $\qca^{s}(\qubit{\xi_n}|\qubit{1^n})\leq d$ for
some absolute constant $d\geq0$.
\end{lemma}

In connection to distinguishability, Sipser \cite{Sip83} defined the
notion of distinguishing complexity, which measures the minimal size
of a program that distinguishes a target classical string from all
other strings. Translating this distinguishing complexity into a
quantum context, we introduce the {\em $k$-separability distinguishing
complexity} of a partially entangled state.

\begin{definition}\label{def:sQCD}
Let $s$ be any function from $\nat$ to $\nat$ and let
$k\in\nat-\{0,1\}$. For any qustring $\qubit{\xi}$ of length $n$, the
{\em $s$-size bounded $k$-separability distinguishing complexity} of
$\qubit{\xi}$, denoted $\sqcd^{s}_{k}(\qubit{\xi})$, is defined to be
the infimum of $\cc(D|k) - \log{\epsilon}$ for any quantum circuit $D$
of size at most $s(n)$ with $n^2/2+7n/2+k+2$ inputs and (possibly)
ancilla qubits such that $D$ $\epsilon$-distinguishes between
$\qubit{1^{\sigma,\vec{m}}}\qubit{\xi}$ and
$\qubit{1^{\sigma,\vec{m}}}\qubit{\phi}$ for any $k$-separable
qustring $\qubit{\phi}$ of length $n$ and any permutation $\sigma$
that achieves the $k$-separability of $\qubit{\phi}$ with
$k$-sectioning $\vec{m}$.  For convenience, we define
$\sqcd^{s}_{\sigma,\vec{m}}(\qubit{\xi})$ similarly by requiring
conditions (i) and (ii) to hold only for the fixed pair
$(\sigma,\vec{m})$. The conditional version
$\sqcd^{s}_{k}(\qubit{\xi}|\qubit{\zeta})$ is defined by
$\cc(D|k,\ell(\qubit{\zeta})) - \log{\epsilon}$, where $D$ takes
$\qubit{1^{\sigma,\vec{m}}}\qubit{\psi}\qubit{\zeta}$ as input.
\end{definition}

It is important to note that if $\qubit{\xi}$ is $k$-separable then
$\sqcd^{s}_{k}(\qubit{\xi})$ is {\em not defined} since $\epsilon$
becomes zero. The next lemma follows immediately from Definition
\ref{def:sQCD}.

\begin{lemma}\label{lemma:sQCD}
Let $k\geq2$ and let $\qubit{\xi}$ be any qustring.

1. $\sqcd^{s}_{\sigma,\vec{m}}(\qubit{\xi}) \leq
\sqcd^{s}_{k}(\qubit{\xi})$ for any permutation $\sigma$ and
$k$-sectioning $\vec{m}$.

2. $\sqcd^{s}_{k+1}(\qubit{\xi})\leq \sqcd^{s}_{k}(\qubit{\xi})$ if
   $k\leq \ell(\qubit{\xi})-1$.

3. Let $k,s$ be any functions from $\nat$ to $\nat$ with $k(n)\geq2$
   for all $n$. Let $\epsilon$ be any function from $\nat$ to
   $(0,1]$. If an ensemble $\Xi=\{\qubit{\xi_n}\}_{n\in\nat}$ is
   uniformly $(k,\epsilon,s)$-distinguishable from separable states,
   then there exists a constant $c\geq0$ such that
   $\sqcd^{s}_{k(n)}(\qubit{\xi_n}|\qubit{1^n})\leq c-
   \log{\epsilon(n)}$ for all $n\in\nat$. In particular, if
   $\epsilon(n)$ is bounded above by a certain constant, then
   $\sqcd^{s}_{k(n)}(\qubit{\xi_n}|\qubit{1^n})\leq d$ for some
   absolute constant $d\geq0$.
\end{lemma}

Note that if $\sqcd^{s}_{k}(\qubit{\xi})=\cc(D|k)-\log{\epsilon}$ as
in Definition \ref{def:sQCD} then $D$ is a
$(k,\epsilon,s)$-distinguisher of $\qubit{\xi}$. By (the proof of)
Proposition \ref{prop:close-indistinguish}, $\epsilon$ cannot be less
than or equal to $sdis_{k}(\qubit{\xi})$. This gives a lower bound of
separability distinguishing complexity.

\begin{proposition}
For any qustring $\qubit{\xi}$ and any integer $k$ with $2\leq k\leq
\ell(\qubit{\xi})$, if $sdis_{k}(\qubit{\xi})>0$, then
$\sqcd^{s}_{k}(\qubit{\xi}) > - \log{sdis_{k}(\qubit{\xi})}$.
\end{proposition}

At length, we exhibit two upper bounds of separability distinguishing
complexity, which follow from Lemma \ref{lemma:construct-distinguish}
and Proposition \ref{prop:approximate-distinguish}. Note that
Proposition \ref{prop:upper-bounds} requires a calculation slightly
different from Proposition \ref{prop:approximate-distinguish}.

\begin{proposition}\label{prop:upper-bounds}
Let $\Xi=\{\qubit{\xi_n}\}_{n\in\nat}$ be any entanglement ensemble
with size factor $\ell$. Let $k,s$ be any functions from $\nat$ to
$\nat$ with $2\leq k(n)\leq \ell(n)$ for all $n$.

1.  If $\Xi$ is $s$-size constructible, then there exist a constant
$c\geq0$ and a function $s'(n)\in O(s(n))$ such that
$\sqcd^{s'}_{k}(\qubit{\xi_n})\leq \qca^{s}(\qubit{\xi_n}) -
2\log{sdis_{k(n)}(\qubit{\xi_n})} +c$ for all $n\in\nat$.

2. If $\Xi$ is $(\epsilon,s)$-approximable and
$sdis_{k(n)}(\qubit{\xi_n})>2\sqrt{\epsilon(n)}$ for all $n$, then
there exist a constant $c\geq0$ and a function $s'(n)\in O(s(n))$ such
that, for all $n$'s, $\sqcd^{s'}_{k}(\qubit{\xi_n})\leq
\qca^{s}(\qubit{\xi_n}) - \log{sdis_{k(n)}(\qubit{\xi_n})} -
\log\left(\frac{sdis_{k(n)}(\qubit{\xi_n})
-2\sqrt{\epsilon(n)}}{1-\epsilon(n)^2}\right) +c$.
\end{proposition}

\bibliographystyle{alpha}

\begin{thebibliography}{Gur91}

\bibitem{BBC+93} 
C. H. Bennett, G. Brassard, C. Cr{\'e}peau, R. Jozsa,
A. Peres, and W. Wootters. Teleporting an unknown quantum state via
dual classical and EPR channels. {\em Phys. Rev. Lett.}, {\bf 70}
(1993), 1895--1899.

\bibitem{BV97}
E. Bernstein and U. Vazirani. Quantum complexity theory. {\em SIAM
J. Comput.}, {\bf 26}, 1411--1473, 1997.

\bibitem{BDL00}
A. Berthiaume, W. van Dam, and S. Laplante. Quantum Kolmogorov
complexity. To appear in {\em J. Comput. System Sci.} 
See also ArXive e-print
quant-ph/0005018, 2000.

\bibitem{BW92}
C. H. Bennett and S. J. Wiesner. Communication via one and
two-particle operations on Einstein-Podolsky-Rosen states. {\em
Phys. Rev. Lett.}, {\bf 69} (1992), 2881--2884.

\bibitem{BCWW01}
H. Buhrman, R. Cleve, J. Watrous, R. de Wolf. Quantum
fingerprinting. {\em Phys. Rev. Lett.}, {\bf 87}:167902 (2001).

\bibitem{Deu89}
D. Deutsch. Quantum computational networks. 
{\em Proc. Roy. Soc. London.} A {\bf 425} (1989), 73--90. 

\bibitem{FG99}
C. A. Fuchs and J. van de Graaf. Cryptographic distinguishability
measures for quantum-mechanical states. {\em IEEE Transactions on
Information Theory}, {\bf 45} (1999), 1216--1227.

\bibitem{Hor01} 
M. Horodecki. Entanglement measures. {\em
Quant. Info. Comp.} {\bf 1} (2001), 3--26.

\bibitem{Kni95}
E. Knill. Approximating quantum circuits. ArXive e-print
quant-ph/9508006, 1995.

\bibitem{KMY01} 
H. Kobayashi, K. Matsumoto, and T. Yamakami. Quantum
Merlin-Arthur proof systems: are multiple Merlins more helpful to
Arthur? In this proceedings.

\bibitem{NC00}
M. A. Nielsen and I. L. Chuang. {\em Quantum Computation and
Information}. Cambridge University Press, 2000.

\bibitem{Sip83}
M. Sipser. A complexity theoretic approach to randomness. In {\em
Proceedings of the 15th ACM Symposium on the Theory of Computing},
pp.330--335, 1983.

\bibitem{VPRK97} 
V. Vedral, M. B. Plenio, M. A. Rippin, and
P. L. Knight. Quantifying entanglement. {\em Phys. Rev. Lett.} {\bf
78} (1997), 2275--2279.

\bibitem{VPJK97} 
V. Vedral, M. B. Plenio, K. Jacob, and
P. L. Knight. Statistical inference, distinguishability of quantum
states, and quantum entanglement. {\em Phys. Rev.} A {\bf 56} (1997),
4452--4455.

\bibitem{Vit01}
P. M. B. Vit{\'a}nyi. Quantum Kolmogorov complexity based on classical
descriptions. {\em IEEE Transactions on Information Theory}, {\bf 47}
(2001), 2464--2479.

\bibitem{WG03} 
T. C. Wei and P. M. Goldbart. Geometric measure of
entanglement and applications to bipartite quantum states. ArXive
e-print quant-ph/0307219, 2003.

\bibitem{Yam99}
T. Yamakami. A foundation of programming a multi-tape quantum Turing
machine. In {\em Proceedings of the 24th International Symposium on
Mathematical Foundations of Computer Science}, Lecture Notes in
Computer Science, Vol.1672, pp.430--441, Springer-Verlag, 1999.

\bibitem{Yam02} 
T. Yamakami. Quantum NP and a quantum hierarchy. In
{\em Proceedings of the 2nd IFIP Conference on Theoretical Computer
Science} (Foundations of Information Technology in the Era of Network
and Mobile Computing), pp.323--336, Kluwer Academic Publishes, 2002.

\bibitem{Yao93}
A. C. Yao. Quantum circuit complexity. In {\em Proceedings of the 34th
Annual Symposium on Foundations of Computer Science}, pp.352--361,
1993.

\end{thebibliography}

\end{document}